\documentclass[letterpaper, 10 pt, conference]{FILES/ieeeconf}
\IEEEoverridecommandlockouts
\usepackage{cite}
\usepackage{amsmath,amssymb,amsfonts}
\usepackage{graphicx}
\usepackage{textcomp}
\usepackage{comment}
\usepackage{hyperref}
\usepackage{array}
\usepackage{ragged2e}
\usepackage{tikz}
\usetikzlibrary{shapes,arrows}
\usepackage[linesnumbered,ruled,vlined]{algorithm2e}
\usepackage{fancyhdr}
\usepackage{amsmath,amssymb,amsfonts}
\usepackage{graphicx}
\usepackage{xcolor}
\usepackage{float}
\usepackage{fancyhdr}
\usepackage{lastpage}
\usepackage{eso-pic}
\usepackage{fancyhdr}
\usepackage{multirow}
\usepackage{flushend}

\usepackage{orcidlink}
\usepackage{subcaption}

\newcolumntype{P}[1]{>{\RaggedRight\arraybackslash}p{#1}}

\def\BibTeX{{\rm B\kern-.05em{\sc i\kern-.025em b}\kern-.08em
    T\kern-.1667em\lower.7ex\hbox{E}\kern-.125emX}}
\fancypagestyle{FirstPage}{
\lfoot{979-8-3503-4536-0/23/\$31.00 \copyright2023 IEEE} 
}



\usepackage{eso-pic}
\newcommand\AtPageUpperMyleft[1]{\AtPageUpperLeft{
 \put(\LenToUnit{0.077\paperwidth},\LenToUnit{-1cm}){
     \parbox{0.9\textwidth}{\raggedright\fontsize{9}{11}\selectfont #1}}
 }}
\newcommand{\conf}[1]{
\AddToShipoutPictureBG*{
\AtPageUpperMyleft{#1}
}
}

\conf{2023 26th International Conference on Computer and Information Technology (ICCIT)\\
  13-15 December, Cox’s Bazar, Bangladesh}

\title{\LARGE \bf 
Attention Based Feature Fusion Network for \\Monkeypox Skin Leison Detection
}

\author{Niloy Kumar Kundu \orcidlink{0009-0007-5520-574X}, Mainul Karim, Sarah Kobir and Dewan Md. Farid
\thanks{Niloy Kumar Kundu, Mainul Karim, Sarah Kobir and Dewan Md. Farid with the Department of Computer Science and Engineering, United International University, United City, Madani Avenue, Badda, Dhaka 1212, Bangladesh
        {\tt\small nkundu191071@bscse.uiu.ac.bd, dewanfarid@cse.uiu.ac.bd}}
}

\begin{document}

\maketitle
\thispagestyle{empty}
\pagestyle{empty}

\begin{abstract}
The recent monkeypox outbreak has raised significant public health concerns due to its rapid spread across multiple countries. Monkeypox can be difficult to distinguish from chickenpox and measles in the early stages because the symptoms of all three diseases are similar. Modern deep learning algorithms can be used to identify diseases, including COVID-19, by analyzing images of the affected areas. In this study, we introduce a lightweight model that merges two pre-trained architectures, EfficientNetV2B3 and ResNet151V2, to classify human monkeypox disease. We have also incorporated the squeeze-and-excitation attention network module to focus on the important parts of the feature maps for classifying the monkeypox images. This attention module provides channels and spatial attention to highlight significant areas within feature maps. We evaluated the effectiveness of our model by extensively testing it on a publicly available Monkeypox Skin Lesions Dataset using a four-fold cross-validation approach. The evaluation metrics of our model were compared with the existing others. Our model achieves a mean validation accuracy of 96.52\%, with precision, recall, and F1-score values of 96.58\%, 96.52\%, and 96.51\%, respectively.
\newline

\indent \textit{Deep Learning; Transfer Learning; Squeeze- and-Excitation Network;}
\end{abstract}

\section{INTRODUCTION}
\label{section1}
Skin cancer has been increasing day by day, particularly in the case of monkeypox. It is caused by the monkeypox virus which can also affect both humans and animals due to its zoonotic nature \cite{al2022review}. In 1970, monkeypox was first identified in humans and the majority of cases are reported in West Africa \cite{durski2018emergence}. The World Health Organization (WHO) designated the worldwide monkeypox epidemic as an international public health emergency on July 23, 2022 \cite{mitja2023monkeypox}. Detecting monkeypox is the most challenging task due to its nonspecific symptoms and the unavailability of a  polymerase chain reaction (PCR) test. Likewise, the early identification of monkeypox becomes challenging because it shares common characteristics with other illnesses like roseola, smallpox, and scarlet fever.

Monkeypox symptoms can be serious and usually last for 2 to 4 weeks. WHO report states that the recent case mortality ratio has been between 3\% and 6\% \cite{who}. In general, the incubation period lasts from 6 to 13 days. The period can vary, often ranging from 5 to 21 days \cite{li2006detection}. Typically, skin rashes first emerge on the face before extending to other body parts. Additionally, areas such as the genital region, eyes, and the mucous membrane inside the mouth may also develop lesions. Patients experience a variety of symptoms during the invasion stage including headache, muscle aches, skin rash, high fever, lymph node swelling, chills, and fatigue \cite{shafaati2022monkeypox}.

Deep learning algorithms have seen widespread application in image processing and pattern recognition in the past few years. Furthermore, it has also been demonstrated that deep learning models can easily identify hidden features and classify various types of diseases \cite{affonso2017deep}. In the medical field, deep learning algorithms are commonly employed these days to uncover concealed attributes via a variety of convolutional processes \cite{shen2017deep}. Different types and variations of convolutional neural networks (CNNs) are employed by researchers to tackle a range of complex classification problems. In addition, Ensemble learning uses multiple classifiers instead of a single one. It combines multiple classifiers to predict the class level of unknown items \cite{rincy2020ensemble}. By integrating various classifiers, the model leverages the strengths of the foundational classifiers, thereby boosting its overall performance.


The method of transfer learning entails the use of a model that has been pre-trained on a substantial dataset (like ImageNet), and applying its learned knowledge to a new task that utilizes a smaller target dataset \cite{yu2022transfer}. This study puts forth an approach based on ensemble learning for the detection of the monkeypox virus from images of skin lesions, utilizing the method of transfer learning. We have incorporated two pre-trained-based learners, namely EfficientNetV2B3 and ResNet152V2, to fine-tune the dataset. We have also incorporated the squeeze-and-excitation network (SE-Net) \cite{hu2018squeeze} block to enhance feature extraction capabilities through attention mechanisms and improve the overall performance of the model. After that, we concatenated the feature maps from the two pre-trained architectures and passed them through two dense blocks. Our model achieved a mean validation accuracy of 96.52\%.

The remainder of the study is organized as follows: Section \ref{methodology} is dedicated to summarizing recent studies on the classification of monkeypox disease using deep learning algorithms. Section \ref{methodology} describes the proposed methodology and related materials required for the study. In Section \ref{results}, the description of datasets, hyper-parameter tuning, k-fold cross-validation technique, evaluation matrices, experimental setup, and experimental results are briefly described. Finally, Section \ref{conclusion_futureWork} concludes the paper.
One Section \ref{methodology} of the paper is devoted to reviewing recent research on the use of deep learning algorithms for monkeypox disease classification. Another Section \ref{methodology} outlines the proposed methodology and the necessary materials for the study. A subsequent Section \ref{results} provides a brief overview of the datasets, hyper-parameter tuning, k-fold cross-validation technique, evaluation metrics, experimental setup, and experimental results. The paper is concluded in the final Section \ref{conclusion_futureWork}.

\section{Related Works}
\label{relatedWorks}

Skin lesions are common in medical practice, and accurate identification is essential for effective patient care \cite{almufareh2023transfer}. In recent times, researchers have focused their efforts on exploring various kinds of deep learning algorithms with the monkeypox disease. This literature review offers a comprehensive analysis of pertinent studies concentrating on the detection of monkeypox skin lesions through the application of deep learning methodologies in image analysis.

Ali et al. \cite{ali2022monkeypox} addressed the issue of limited data to train the model for classifying the monkeypox skin lesions. To address this gap, they collected images from websites, news portals, and public case reports and introduced a monkeypox skin lesion dataset. They also performed data augmentation to increase the number of data. Then classification was performed using several pre-trained models, such as VGG16, ResNet50, Inception V3, and an ensemble of these three pre-trained architectures on their self-created MLSD dataset. Among them, ResNet50 achieves a height accuracy of 82.96±(4.57)\%.

Haque et al. \cite{haque2023ensemble} proposed a model that takes five pre-trained models, namely ResNet50, ResNet152V2, DenseNet121, InceptionV3, and EfficientNetV2B3. They trained each model using two different datasets, including “Monkeypox-dataset-2022 (MD-2022)” and “Monkeypox Skin Images Dataset (MSID)”. After that, they combine the models using the majority voting approach. The ensemble model achieves an accuracy of 98.7\% on the MSID dataset and 89.4\% on the MD-2022 dataset.

Almufareh et al. \cite{almufareh2023transfer} perform four different pre-trained models including Inception V3, ResNet50, MobileNet V2, and EfficientNet-B4 across two different datasets. On the Monkeypox Skin Image Dataset (MSID) dataset, the proposed model achieved 93\% accuracy. 

In addition, Sahin et al. \cite{sahin2022human} expanded their research to practical applications by creating a website and also designing an Android app using deep learning models for classifying monkeypox diseases. The Android app captures monkeypox images using the Camera2 API on the Android platform and transmits them to the pre-trained models for analysis. Based on the MSLD dataset, they highlighted the performance of MobileNetV2, which achieves an accuracy of 91.11\% for binary class classification among the six pre-trained models.

Haque et al. \cite{haque2022classification} implemented five different models including VGG19, Xception, DenseNet121, EfficientNetB3, and MobileNetV2 along with a convolutional block attention module (CBAM). Among these five different pre-trained architectures Xeption-CBAM-Dense architecture performs well and the model achieved 83.89\% validation accuracy.

Irmak et al. \cite{irmak2022monkeypox} performed a classification task using pre-trained CNN models on the MISD including MobileNetV2, VGG16, and VGG19. Among these models, MobileNetV2 achieved the highest accuracy rate at 91.37\%, accompanied by 90.5\% precision, 86.75\% recall, and an F1-score of 88.25\%.

Altun \cite{altun2023monkeypox} implemented a convolutional neural network (CNN) model using pre-trained models, such as ResNet50, EfficientNetV2, MobileNetV3-s, Vgg19, DenseNet121, and Xception. For the evaluation of the model, they use the AUC, accuracy, recall, loss, and F1-score. Among all the architectures, the MobileNetV3-s model achieves a height accuracy of 97\%. The F1-score of the model is 98\%, with an AUC of 99\%, and a recall of 97\%. 

\section{Methodology}
\label{methodology}

\begin{figure}[htp]
\centering
  \includegraphics[width=9cm]{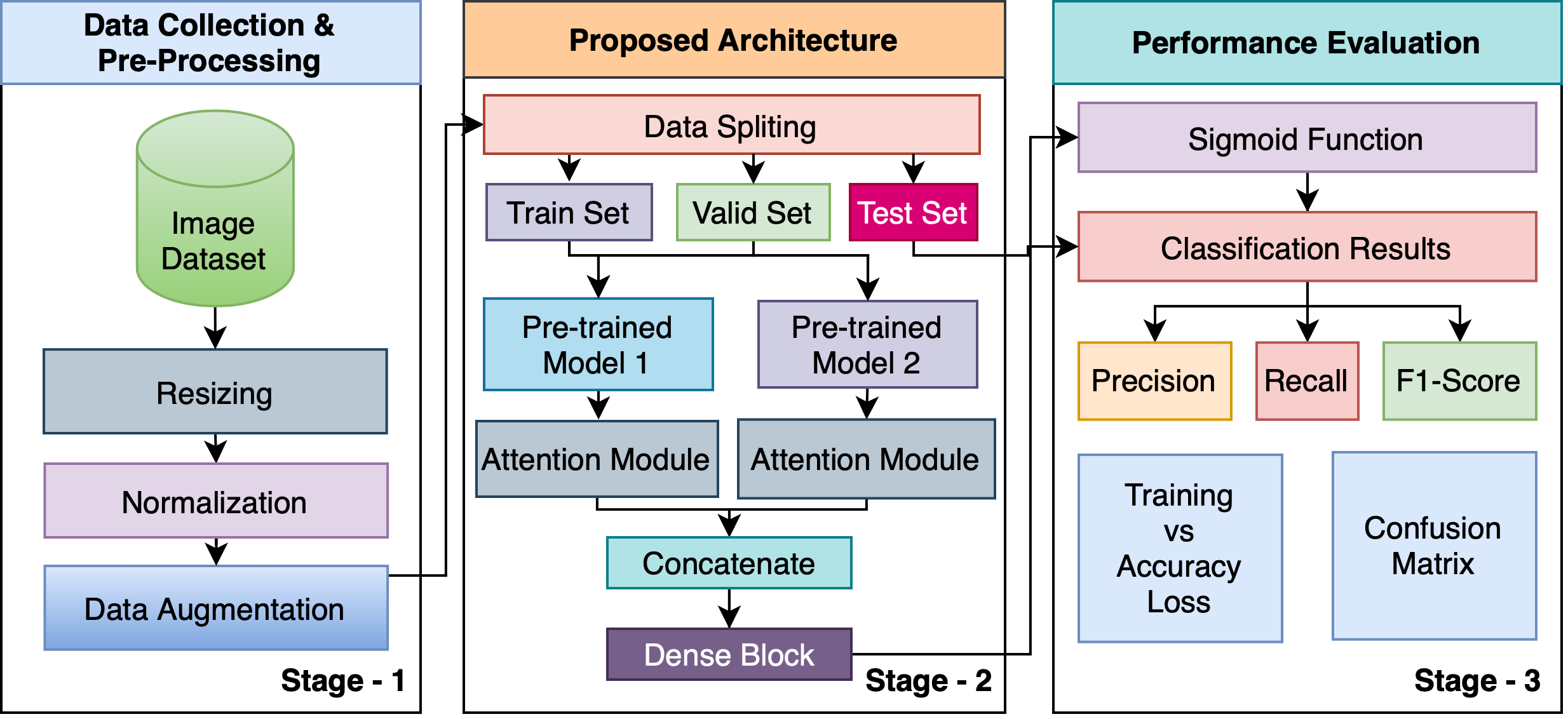}
  \caption{The overall pipeline of our proposed work.}
  \label{fig:workflow}
\end{figure}

The complete workflow of our approach is illustrated in Figure \ref{fig:workflow}. Stage 1 involved data preprocessing of human monkeypox lesion images to prepare them for training. This included tasks such as image resizing, normalization, and augmentation. The images have been resized to a resolution of $224 \times 224 \times 3$. In addition, we have normalized the pixel values from [0-255]. Due to the limitations of the data, data augmentation methods have been performed for better classification results. In stage 2, the dataset has been split into training, validation, and testing sets. Training and validation sets are used to train the proposed model, and using the test set, the overall performance of the model has been evaluated in stage 3.

\begin{figure}[htp]
\centering
  \includegraphics[width=6cm, height=14cm]{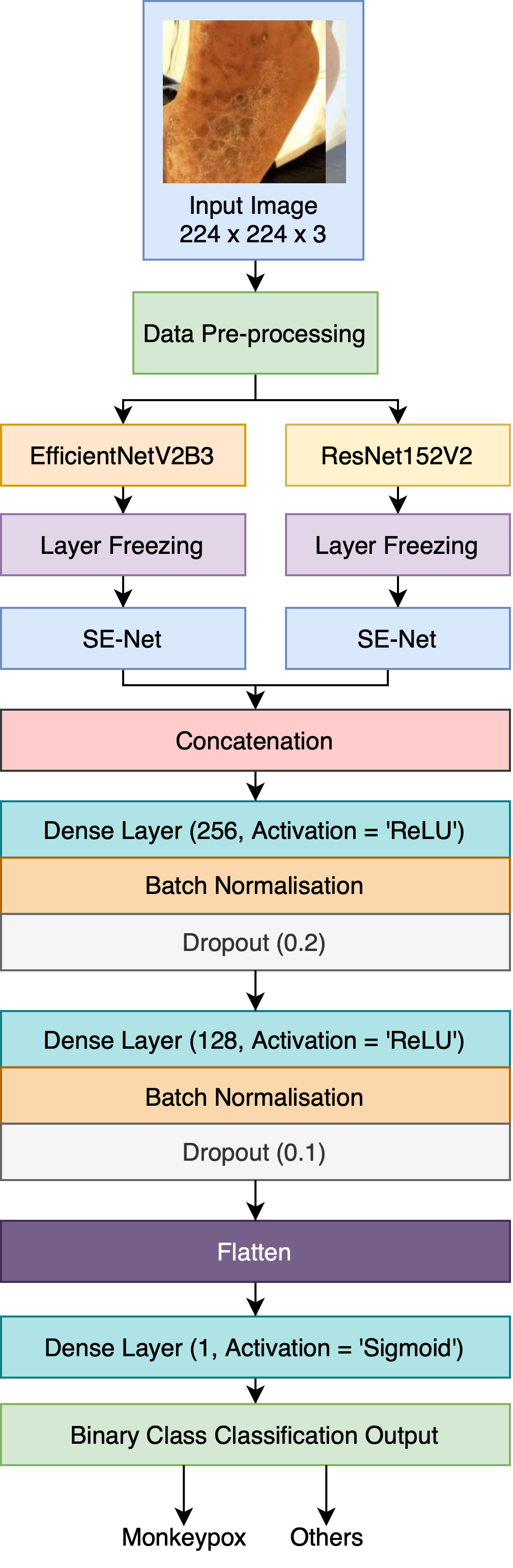}
  \caption{Architectural diagram of our proposed method.}
  \label{fig:model}
\end{figure}

Figure \ref{fig:model} illustrates the complete architectural diagram. We have incorporated EfficientNetV2B3 and ResNet152V2 to extract image features. Additionally, we have frozen all the layers of both of the pre-trained models except the last three layers. EfficientNetV2B3 uses compound scaling to increase the size and depth of the network. So it can capture more hierarchical features from the images. This improves the network's performance in classifying images without overfitting. In ResNet152V2, residual connections are incorporated between each of the convolutional layers. This mechanism enables gradients to flow more smoothly, which helps the network to learn more effectively. Moreover, the SE-Net block offers both channel and spatial attention, concentrating on the most crucial features. The channel attention is directed towards the significant channels, while the spatial attention primarily targets the important spatial locations in the feature maps. Subsequently, the outputs from both SE-Net are combined and fed into two dense blocks. These blocks consist of dense layers with 256 and 128 neurons respectively, and utilize the ReLU activation function. After that, it passes through the batch normalization layers and incorporates dropout layers with rates of 0.2 for the first dense block and 0.1 for the second dense block. After that, we flatten the layers and fed them to the final dense layer with a sigmoid activation function for performing binary class classification. Eq. \eqref{sigmoid} represents the sigmoid activation function.

\begin{equation}
\sigma(x) = \frac{1}{1 + e^{-x}}
\label{sigmoid}
\end{equation}

\subsection{EfficientNetV2B3 }
EfficientNetV2B3 belongs to a family of EfficientNetV2 Architecture \cite{nematzadeh2015comparative}. Moreover, It is based on the EfficientNet architecture and incorporates several modifications to enhance efficiency and accelerate training speed. Compared to other conventional models, EfficientNetV2B3 is designed with specific parameters for scaling, depth, and width to ensure accuracy and efficiency.

\subsection{ResNet152V2}
ResNet152V2 has a deeper network than ResNet-152, which allows it to learn more specific features and understand images better. \cite{he2016deep}. This deep neural network consists of 152 layers. It uses skip connections to solve the vanishing gradient problem. Compared to the original ResNet, ResNetV2 introduces two major changes. The first one is that it applies batch normalization before every weight layer, and the second one is that it uses a bottleneck architecture that shrinks the overall number of parameters.

\subsection{Squeeze-and-Excitation Network (SE-Net)}
\begin{figure}[htp]
\centering
  \includegraphics[width=6cm]{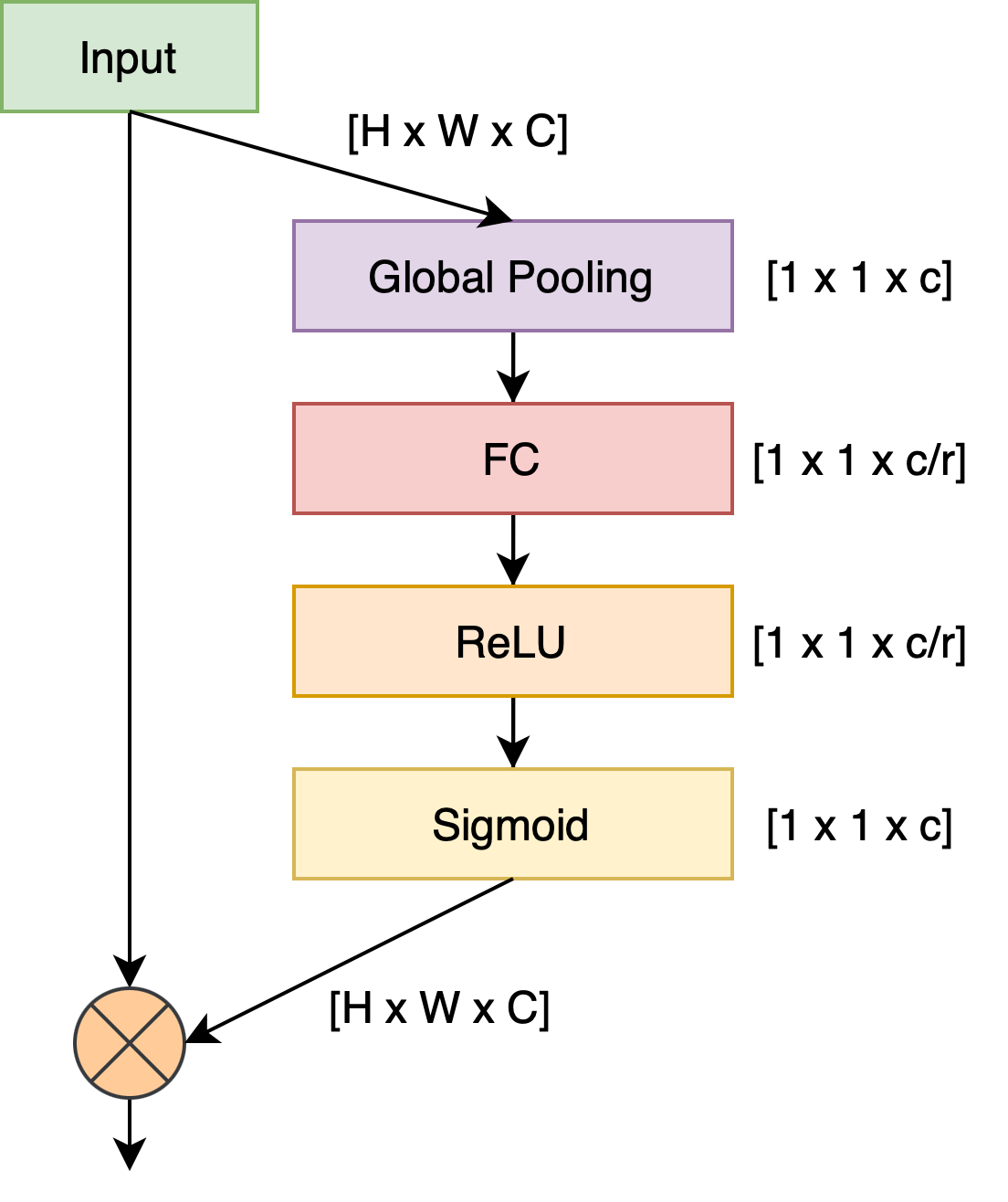}
  \caption{Squeeze-and-Excitation Network (SE-Net) Module \cite{hu2018squeeze}}
  \label{fig:se-net}
\end{figure}

Squeeze-and-Excitation Network (SE-Net) is a technique that focuses on the important features of the data while training \cite{hu2018squeeze}. It involves two steps: squeezing to reduce the feature size and exciting to amplify crucial features based on their importance. For a feature map $X$ with dimensions $C \times H \times W$, the squeeze operation through the global average pooling layer, $Z$ is computed as shown in Eq. \eqref{eq:squeeze}. Fig. \ref{fig:se-net} represents the SE-Net module.

\begin{equation}
\label{eq:squeeze}
Z_c = \frac{1}{H \times W} \sum_{i=1}^{H} \sum_{j=1}^{W} X_{cij}
\end{equation}

Where $Z_c$ is the channel-wise squeeze value for channel c.

The excitation step applies the learnable parameter to apply channel-wise attention to focus on the important features. The learnable parameters are the sigmoid and the ReLU activation functions, respectively. Eq. \eqref{eq:excitation} shows the activation operation.

\begin{equation}
\label{eq:excitation}
E_c = \sigma(W_2 \delta(W_1 Z_c))
\end{equation}

Where $W_1$ and $W_2$ are the weights, $\delta$ represents the ReLU, and $\sigma$ is the sigmoid activation function.
Finally, the excitation values $E_c$ are applied by the original feature map $X$ by element-wise multiplication, as described in Eq. \eqref{eq:final}.

\begin{equation}
\label{eq:final}
Y_{cij} = E_c \cdot X_{cij}
\end{equation}

Through this process, the network can easily identify important features and dynamically adjust their weights to enhance the model’s performance.

\section{Results and Discussion}
\label{results}

\subsection{Dataset Description}
\begin{figure}[ht!]
    \centering
    \begin{subfigure}{0.23\textwidth}
        \includegraphics[width=4cm, height=3cm]{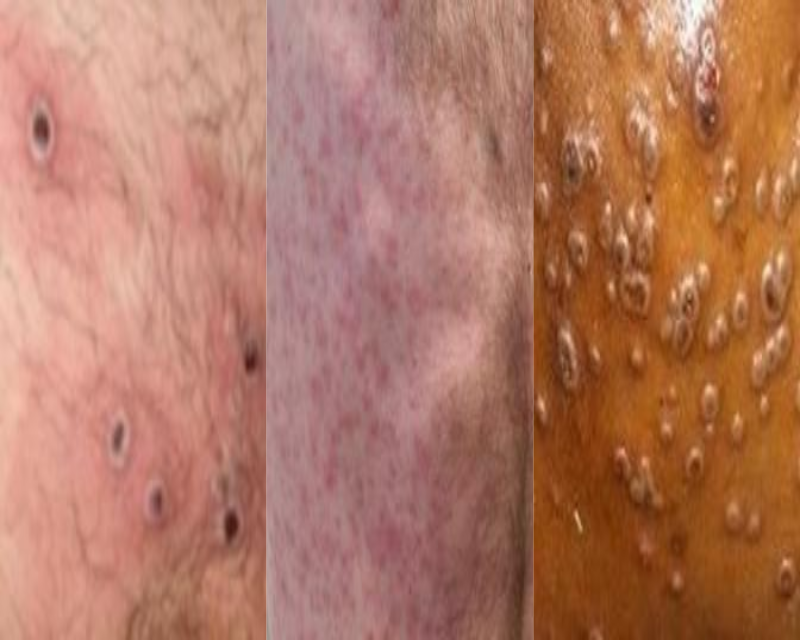}
        \caption{Monkeypox}
        \label{Monkeypox}
    \end{subfigure}
    \hfill
    \begin{subfigure}{0.23\textwidth}
        \includegraphics[width=4cm, height=3cm]{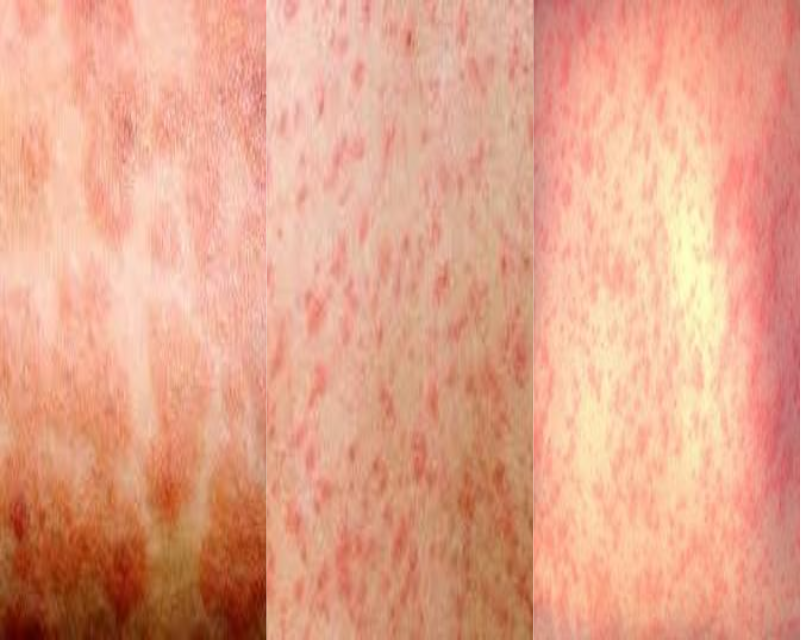}
        \caption{Others}
        \label{Others}
    \end{subfigure}

    \caption{Sample images from the Monkeypox Skin Lesion Dataset.}
    \label{dataset}
\end{figure}

For evaluating our experiment, we used the publicly available Monkeypox Skin Lesion Dataset (MSLD) \cite{ali2022monkeypox}. The dataset is taken from the Kaggle website. Figure \ref{dataset} illustrates the images belonging to two classes within the MSLD dataset.

This dataset contains images of skin lesions that are categorized as either monkeypox or non-monkeypox. Identifying monkeypox can be tricky because it shares symptoms with non-monkeypox diseases like chickenpox and measles, especially in the beginning. The dataset contains three folders, original images, augmented images, and Fold1. Several augmentation methods have been performed, such as reflection, noise, rotation, hue, brightness jitter, saturation, translation, shear, contrast and scaling etc. The information of the original and augmented images are shown in Table \ref{dataset_table}. For our experiment, we used the augmented images folder, which was divided into training, validation, and test sets with a split ratio of 70:20:10.

\begin{table}[htp]
\centering
\caption{Representation of the distribution of both original and augmented images from the Monkeypox Skin Leison Dataset.}
\label{dataset_table}
\begin{tabular}{|c|c|c|}
\hline
\textbf{Type of Image} & \textbf{Original Images} & \textbf{Augmented Images} \\ \hline
\textbf{Monkeypox}     & 102                      & 1428                      \\ \hline
\textbf{Others}        & 126                      & 1764                      \\ \hline
\textbf{Total Images}  & 228                      & 3192                      \\ \hline
\end{tabular}
\end{table}

\subsection{Hyper-parameter tuning}

As a hyperparameter tuner, we chose the Keras Tuner library \cite{omalley2019kerastuner}. This library provides a simple API for automatically tuning hyperparameters to enhance the overall performance of the deep learning model. Table \ref{tuning_block} displays the hyperparameters used for fine-tuning our model. The results of hyperparameter tuning indicate that the initial dense block performed well with 256 neurons in its first dense layer and a dropout rate of 0.2. Similarly, in the second dense block, 128 neurons showed favorable outcomes for the dense layer, and the dropout rate was 0.1. Our model performed the binary classification task using the resulting output parameters. The Adam optimizer was employed to minimize the loss function with a learning rate of 0.0001. We chose a batch size of 32.

\begin{table}[htp]
\centering
\caption{Hyper Parameter tuning}
\label{tuning_block}
\begin{tabular}{|c|c|c|}
\hline
\textbf{Block Name} & \textbf{Dense Layer}   & \textbf{Dropout Layer} \\ \hline
Dense Block - I     & {[}32, 64, 128, 256{]} & {[}0.1, 0.2{]}         \\ \hline
Dense Block - II    & {[}32, 64, 128, 256{]} & {[}0.1, 0.2{]}         \\ \hline
\end{tabular}
\end{table}

\subsection{K-Fold Cross-Validation}
K-fold cross-validation enhances the reliability of the model’s performance \cite{nematzadeh2015comparative}. The dataset is divided into k subsets, where k-1 is used for training and the remaining are used for validation. The process is then repeated k times. In our study, we used a 4-fold cross-validation approach and took the average of these test results across the four folds. Eq. \ref{cross_valid_eq} represents the mathematical equation of the k-fold cross-validation.

\begin{equation}
\label{cross_valid_eq}
\text{K-fold Cross Validation} = \frac{1}{k} \sum_{i=1}^{k} \text{Performance Metric}_i
\end{equation}

\subsection{Performance Metrics}
The results of our proposed model were evaluated using accuracy, precision, recall, and f1-score. The mathematical equation of the matrices is given below.

\begin{equation}
\text{Accuracy} = \frac{TP + TN}{TP + TN + FP + FN}
\end{equation}

\begin{equation}
\text{Precision} = \frac{TP}{TP + FP}
\end{equation}

\begin{equation}
\text{Recall} = \frac{TP}{TP + FN}
\end{equation}

\begin{equation}
\text{F1-Score} = 2 \cdot \frac{\text{Precision} \cdot \text{Recall}}{\text{Precision} + \text{Recall}}
\end{equation}

\subsection{Experimental Setup}
Our model was trained using an NVIDIA Tesla P100 GPU with 16 GB of RAM and the TensorFlow backend. The model is considered lightweight with a total of 73.89 million trainable parameters. For our loss function, we have used binary cross-entropy, as shown in Eq. \eqref{binary}.

\begin{equation}
L(y, \hat{y}) = - \left( y \cdot \log(\hat{y}) + (1 - y) \cdot \log(1 - \hat{y}) \right)
\label{binary}
\end{equation}

\subsection{Experimental Results}
To perform a more accurate classification, we have combined the SE-Net attention mechanism with two pre-trained architectures, EfficientNetV2B3 and ResNet152V2. We have also incorporated dense layers after merging them. The model's performance was evaluated using 4-fold cross-validation to ensure its generalizability and obtain more reliable performance estimates for this classification task. Table \ref{cross-validation} represents the 4-fold cross-validation results.

\begin{table}[htp]
\centering
\caption{Analysis of 4-fold cross-validation}
\label{cross-validation}
\resizebox{0.4\textwidth}{!}{%
\begin{tabular}{|c|c|c|c|c|}
\hline
\textbf{Fold} & \textbf{Accuracy (\%)} & \textbf{Precision (\%)} & \textbf{Recall (\%)} & \textbf{F1-Score (\%)} \\ \hline
1 & 96.87 & 96.90 & 96.87 & 96.87 \\ \hline
2 & 95.65 & 95.70 & 95.65 & 95.64 \\ \hline
3 & 96.70 & 96.78 & 96.70 & 96.69 \\ \hline
4 & 96.87 & 96.94 & 96.87 & 96.86 \\ \hline
\textbf{Mean}   & \textbf{96.52}         & \textbf{96.58}          & \textbf{96.52}       & \textbf{96.51}         \\ \hline
\end{tabular}%
}
\end{table}

We have measured the performance using the accuracy, precision, recall, and f1-score matrices. The result shows that our model achieved 95.65\% mean validation accuracy. For the test data, our model achieves an accuracy of 97.19\%, along with a precision of 97.29\%, a recall of 97.19\%, and an F1-score of 97.19\%. Table \ref{comparison} compares the performance of existing models on the MSLD dataset, using their validation data.

\begin{table}[ht!]
\centering
\caption{A comparative summary of the evaluation metrics obtained from the validation data to compare with other studies that use the same MSLD dataset.}
\label{comparison}
\resizebox{0.48\textwidth}{!}{%
\begin{tabular}{|c|c|c|c|c|}
\hline
\textbf{Model}                                                               & \textbf{Accuracy (\%)} & \textbf{Precision (\%)} & \textbf{Recall} (\%) & \textbf{F1-Score (\%)} \\ \hline
VGG16 \cite{ali2022monkeypox}                      & 81.48         & 85.00          & 81.00       & 83.00         \\ \hline
ResNet50 \cite{ali2022monkeypox}                   & 82.96         & 87.00          & 83.00       & 84.00         \\ \hline
InceptionV3 \cite{ali2022monkeypox}                & 74.07         & 74.00          & 81.00       & 78.00         \\ \hline
Ensemble \cite{ali2022monkeypox}                   & 79.26         & 84.00          & 79.00       & 81.00         \\ \hline
Xception-CBAM-Dense \cite{haque2022classification} & 83.89         & 90.70          & 89.10       & 90.11         \\ \hline
\multirow{2}{*}{\begin{tabular}[c]{@{}c@{}}VGG16 Siamese \\ network model \cite{liu2023convolutional}\end{tabular}} &
  \multirow{2}{*}{88.23} &
  \multirow{2}{*}{90.04} &
  \multirow{2}{*}{86.82} &
  \multirow{2}{*}{88.40} \\
                                                                    &               &                &             &               \\ \hline
MobileNetv2 \cite{irmak2022monkeypox}              & 91.37         & 90.50          & 86.75       & 88.25         \\ \hline
Ensemble \cite{pramanik2023monkeypox}                                                            & 93.39         & 88.91          & 96.78       & 92.35         \\ \hline
\textbf{Proposed Method}                                                     & \textbf{96.52}         & \textbf{96.58}          & \textbf{96.52}       & \textbf{96.51}         \\ \hline
\end{tabular}%
}
\end{table}

\begin{figure}[ht!]
    \centering
    \begin{subfigure}{0.23\textwidth}
        \includegraphics[width=4.3cm]{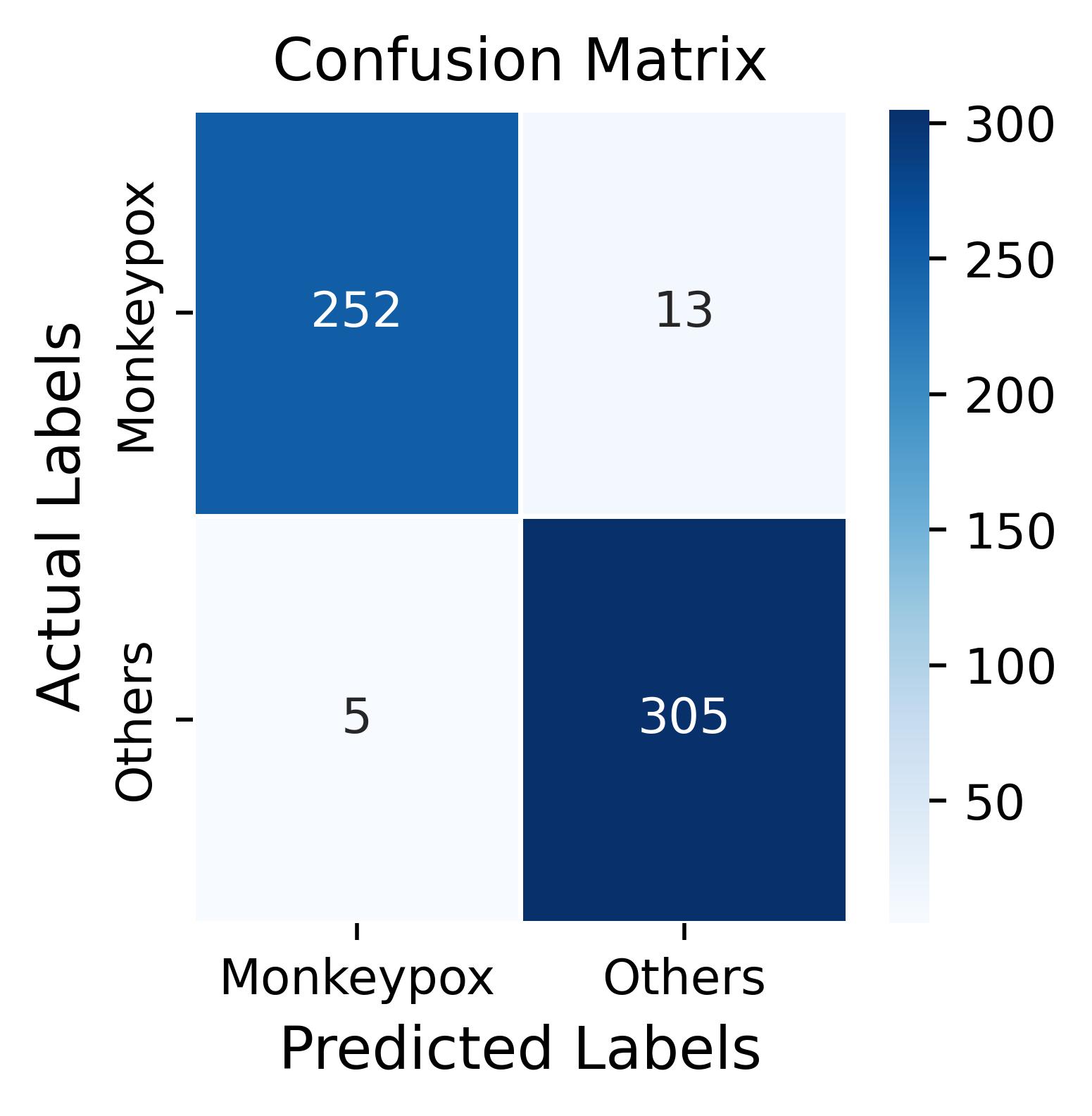}
        \caption{Fold 1}
        \label{fig:accuracy}
    \end{subfigure}
    \hfill
    \begin{subfigure}{0.23\textwidth}
        \includegraphics[width=4.3cm]{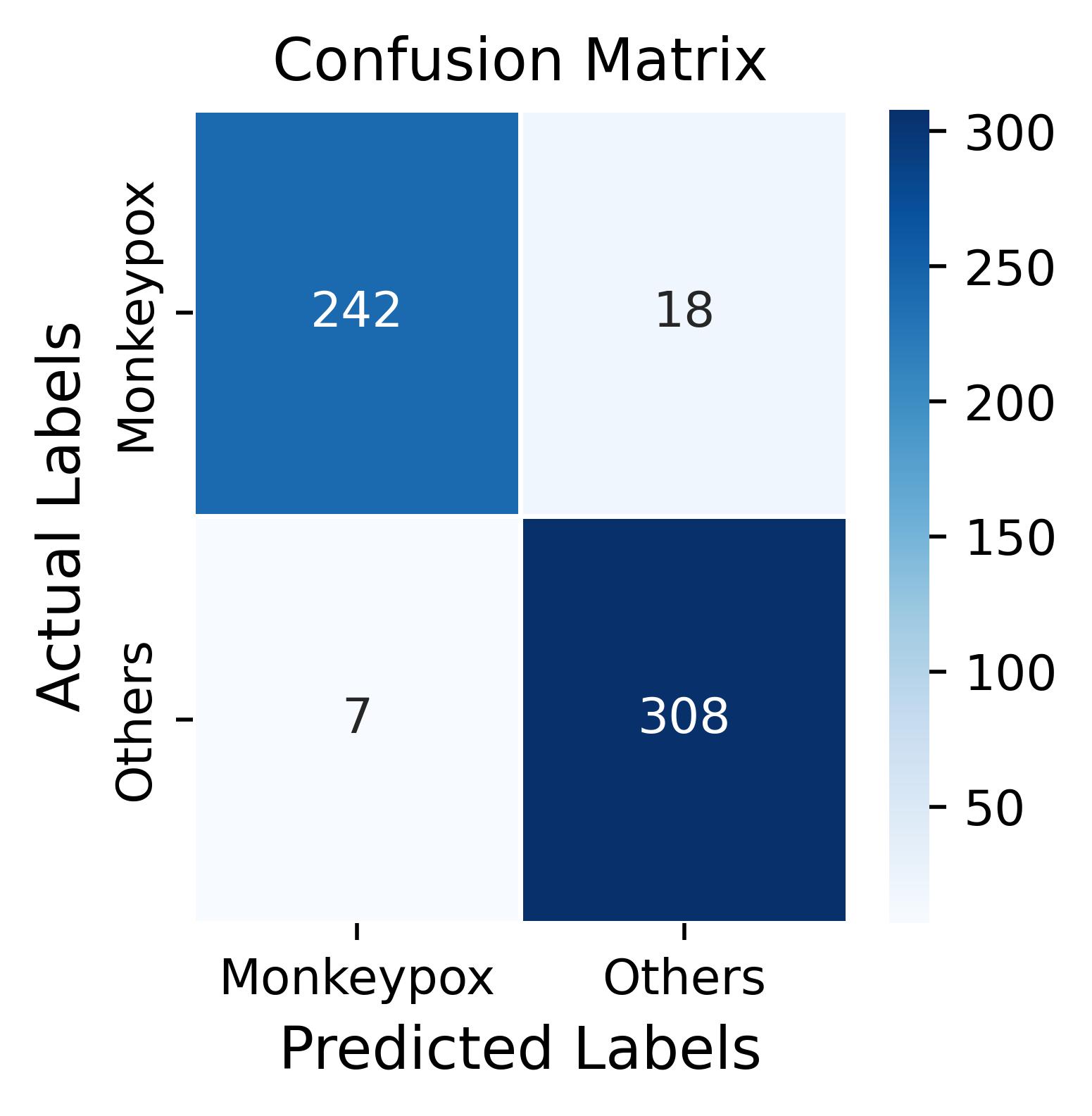}
        \caption{Fold 2}
        \label{fig:loss}
    \end{subfigure}
    \hfill
    \begin{subfigure}{0.23\textwidth}
        \includegraphics[width=4.3cm]{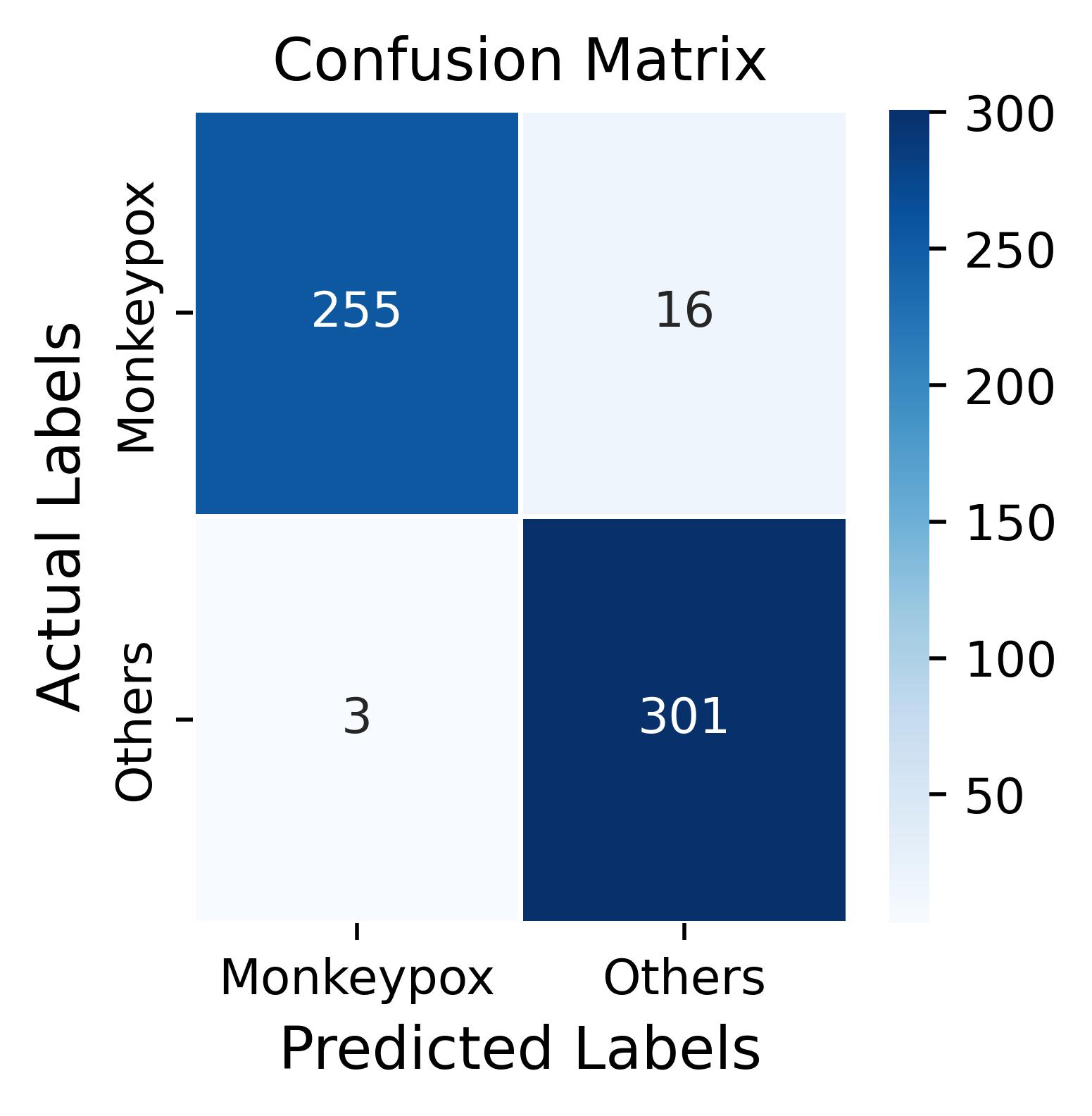}
        \caption{Fold 3}
        \label{fig:loss}
    \end{subfigure}
    \hfill
    \begin{subfigure}{0.23\textwidth}
        \includegraphics[width=4.3cm]{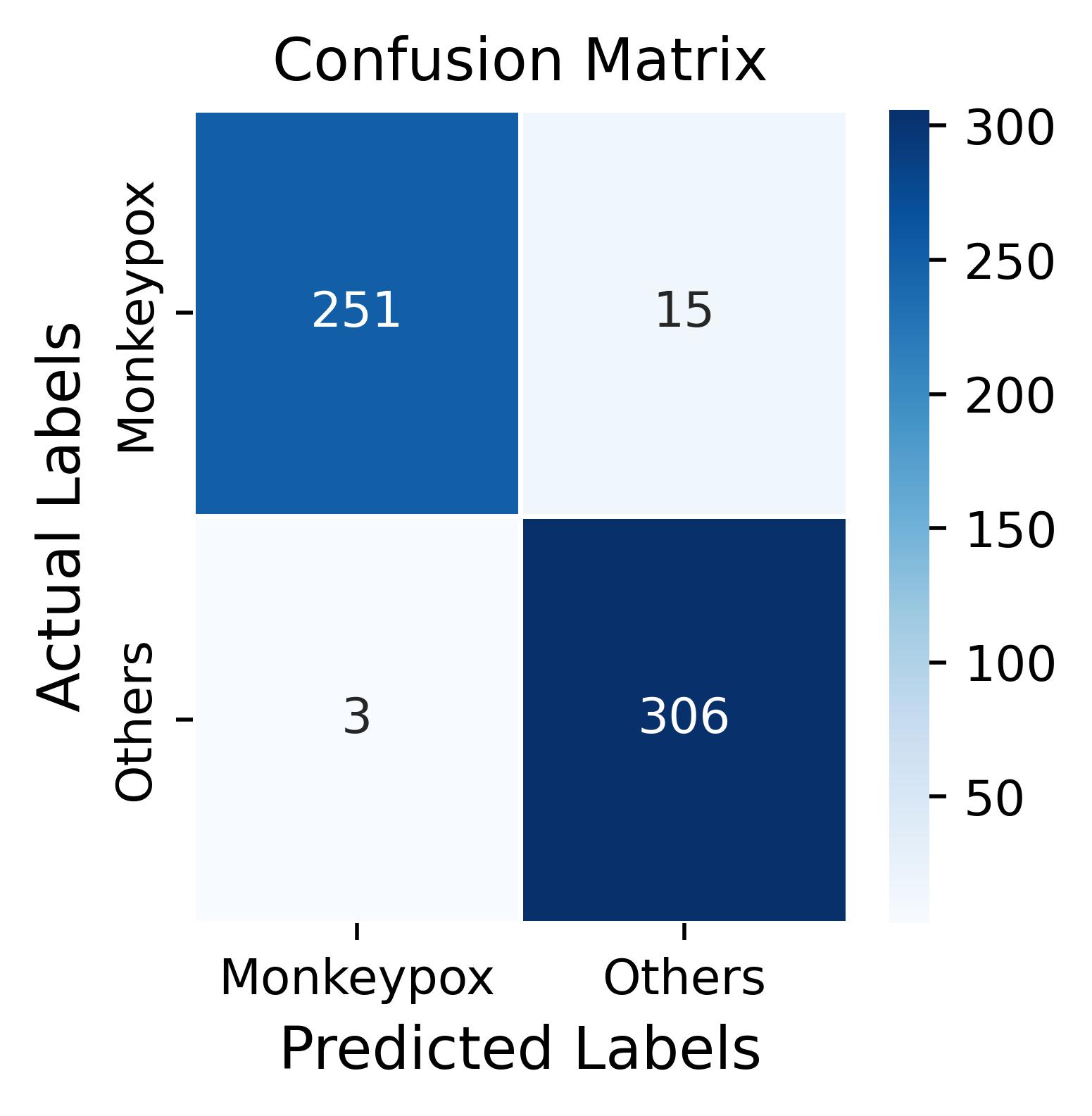}
        \caption{Fold 4}
        \label{fig:loss}
    \end{subfigure}

    \caption{Confusion matrix from the 4-fold cross-validation.}
    \label{confusion}
\end{figure}

Both the ensemble architecture \cite{pramanik2023monkeypox} and our proposed model performs well for classifying the binary classification problem. In addition, our proposed model outperforms all the models based on the given matrices. Figure \ref{confusion} represents the confusion matrix generated by our proposed model.

\begin{figure}[ht!]
    \centering
    \begin{subfigure}{0.23\textwidth}
        \includegraphics[width=4.3cm]{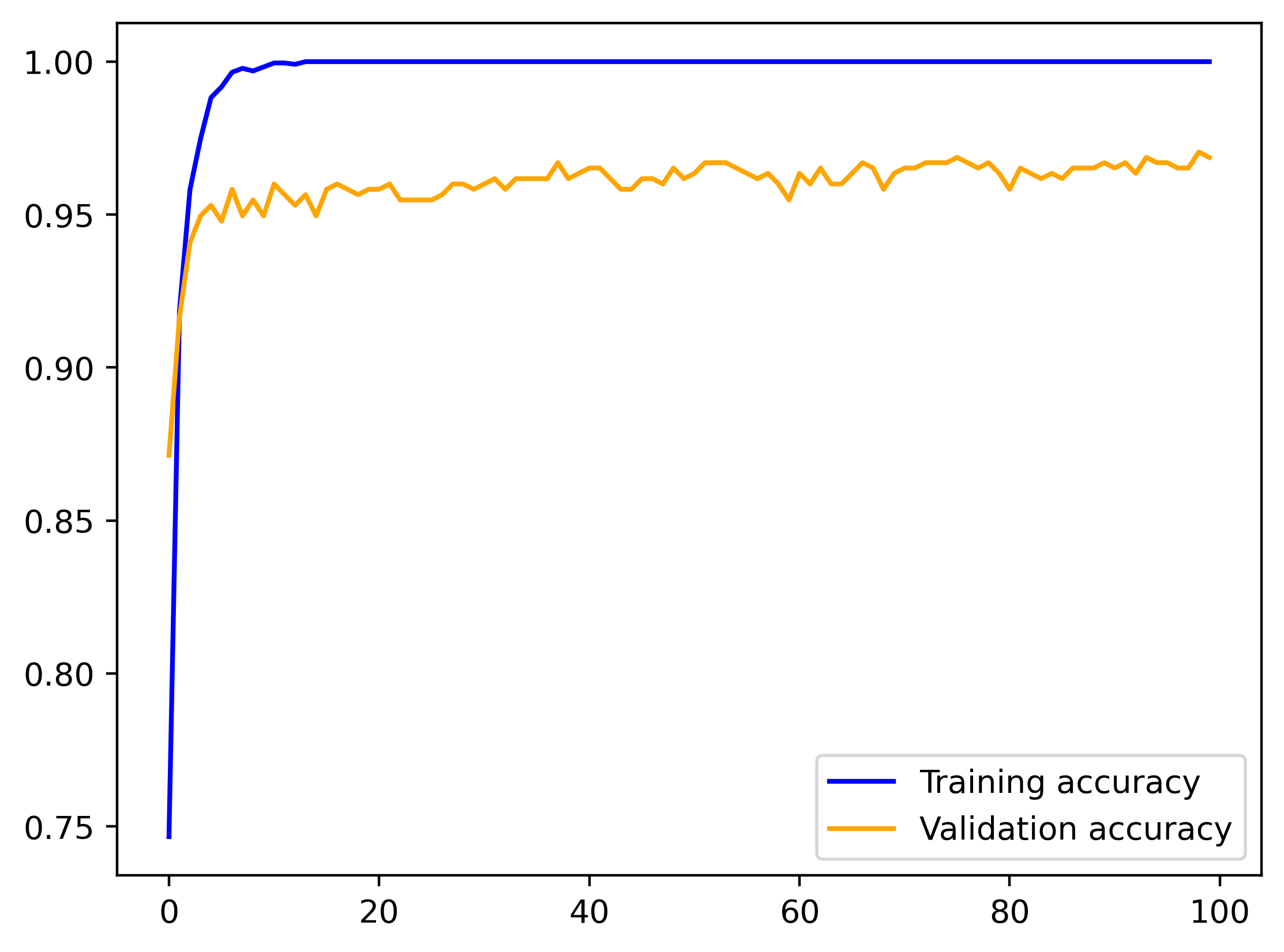}
        \caption{Fold 1}
        \label{fig:accuracy}
    \end{subfigure}
    \hfill
    \begin{subfigure}{0.23\textwidth}
        \includegraphics[width=4.3cm]{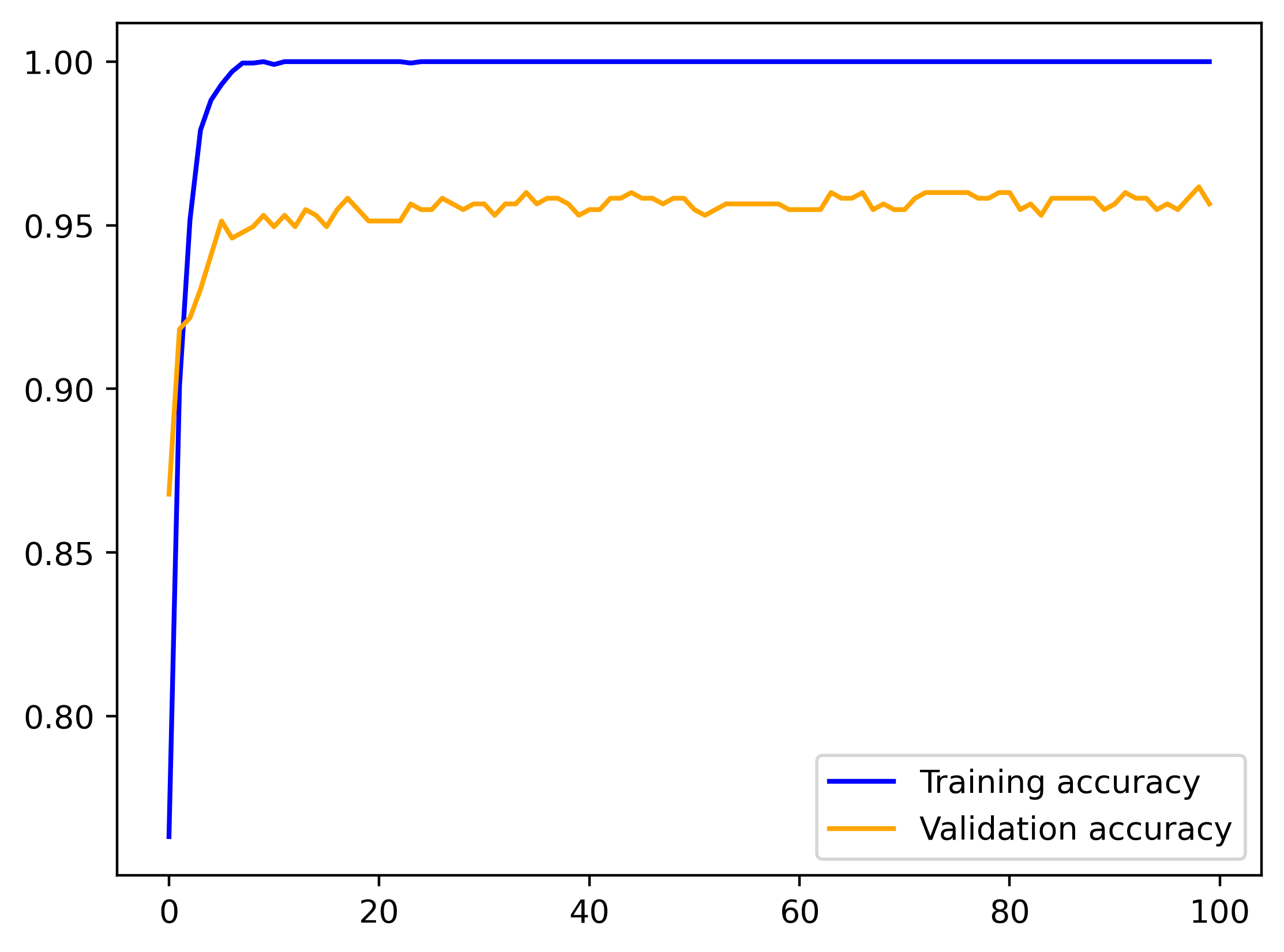}
        \caption{Fold 2}
        \label{fig:loss}
    \end{subfigure}
    \hfill
    \begin{subfigure}{0.23\textwidth}
        \includegraphics[width=4.3cm]{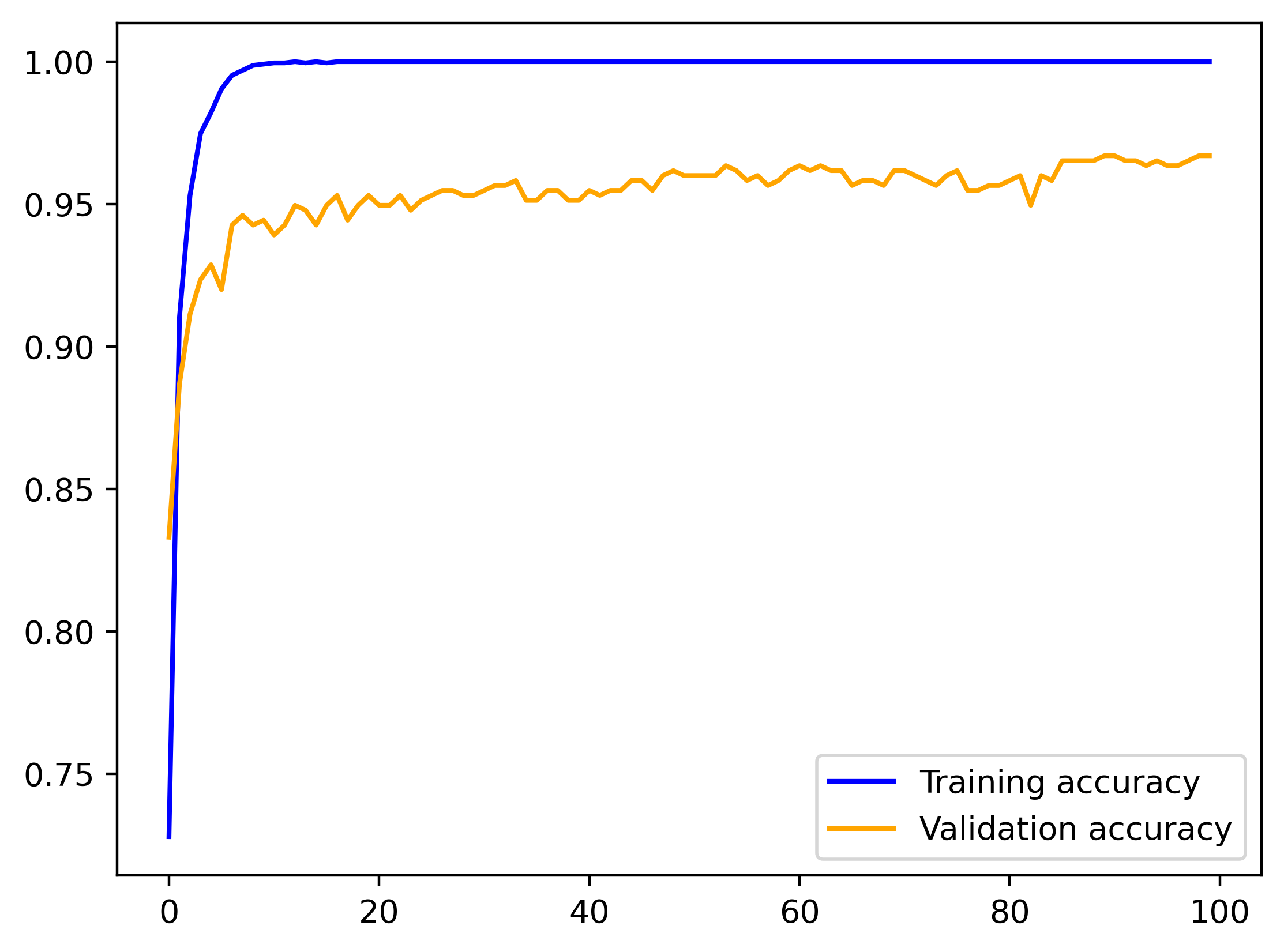}
        \caption{Fold 3}
        \label{fig:loss}
    \end{subfigure}
    \hfill
    \begin{subfigure}{0.23\textwidth}
        \includegraphics[width=4.3cm]{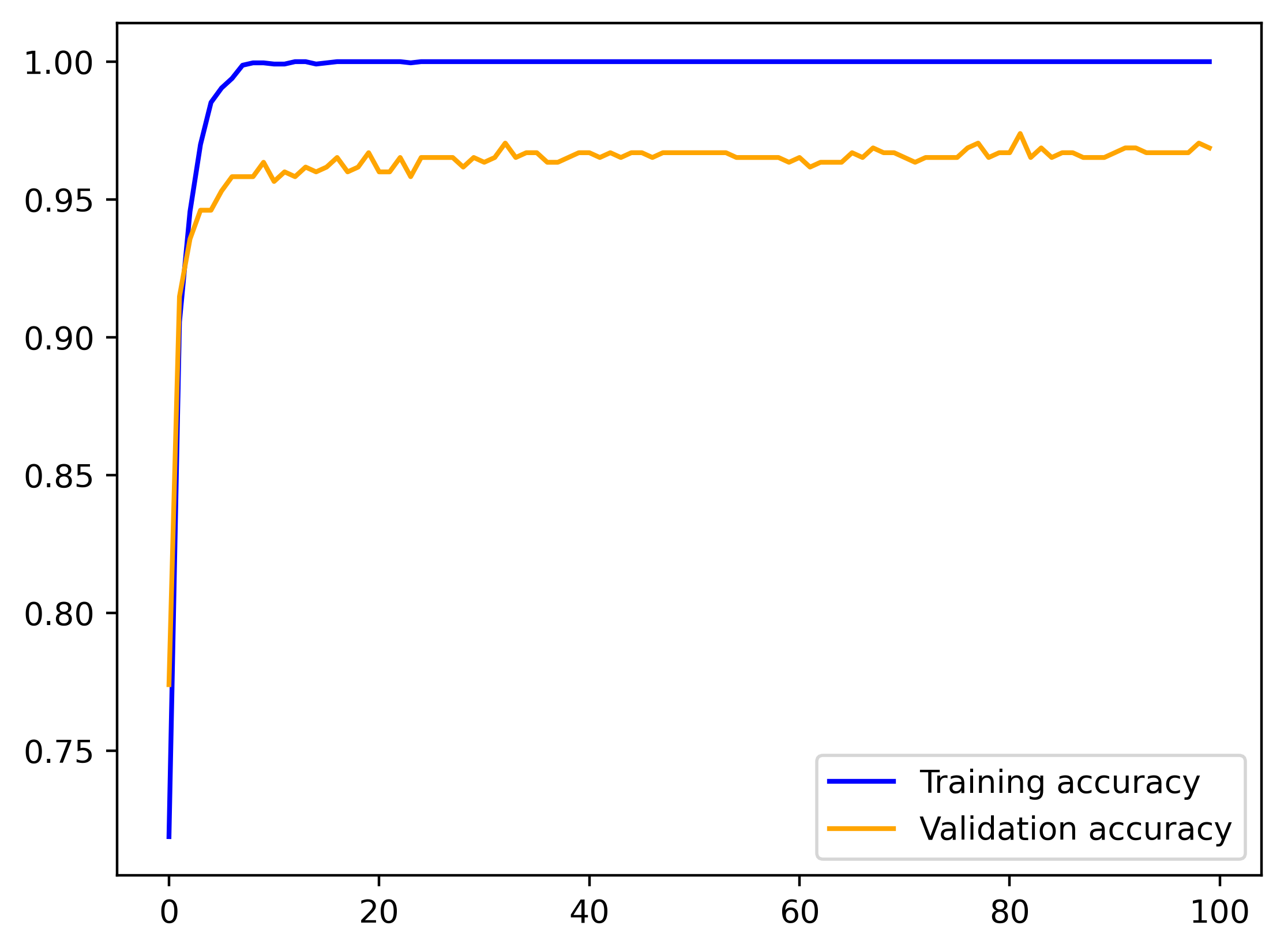}
        \caption{Fold 4}
        \label{fig:loss}
    \end{subfigure}

    \caption{Training vs Validation graph from the 4-fold cross-validation.}
    \label{fig:accuracy_loss}
\end{figure}

Figure \ref{fig:accuracy_loss} represents the training vs. accuracy graphs from the 4-fold cross-validation of our proposed architecture. The graphs demonstrate that both training and validation accuracy have significantly converged.

\section{Conclusion and Future Work}
\label{conclusion_futureWork}
In conclusion, our study introduces a lightweight model for classifying monkeypox diseases. We combine the pre-trained architectures of EfficientNetV2B3 and ResNet152V2, incorporating the SE-Net attention module, and applying dense layers. We conducted the classification task using the MSLD. EfficientNetV2B3 captures the hierarchical features from the images, whereas ResNet152V2 effectively addresses the vanishing gradient problem that occurs in very deep networks. Additionally, the attention mechanism module focuses on channel-wise relationships in feature maps and focuses on the important features. It helps the network to adjust the importance of different channels dynamically during the learning process. The model achieves a mean validation accuracy of 96.52\%, showcasing superior performance in comparison to previous studies utilizing the same dataset. Our study focuses on detecting monkeypox diseases based on 2D images. Future work could explore the use of various deep learning models and possibly extend this approach to include 3D images as well.

\bibliographystyle{FILES/IEEEtran}
\bibliography{main.bib}

\begin{thebibliography}{10}
\providecommand{\url}[1]{#1}
\csname url@samestyle\endcsname
\providecommand{\newblock}{\relax}
\providecommand{\bibinfo}[2]{#2}
\providecommand{\BIBentrySTDinterwordspacing}{\spaceskip=0pt\relax}
\providecommand{\BIBentryALTinterwordstretchfactor}{4}
\providecommand{\BIBentryALTinterwordspacing}{\spaceskip=\fontdimen2\font plus
\BIBentryALTinterwordstretchfactor\fontdimen3\font minus \fontdimen4\font\relax}
\providecommand{\BIBforeignlanguage}[2]{{%
\expandafter\ifx\csname l@#1\endcsname\relax
\typeout{** WARNING: IEEEtran.bst: No hyphenation pattern has been}%
\typeout{** loaded for the language `#1'. Using the pattern for}%
\typeout{** the default language instead.}%
\else
\language=\csname l@#1\endcsname
\fi
#2}}
\providecommand{\BIBdecl}{\relax}
\BIBdecl

\bibitem{al2022review}
S.~Al-Gburi and Z.~Namuq, ``A review of the recent monkeypox outbreak in 2022,'' \emph{Cureus}, vol.~14, no.~8, 2022.

\bibitem{durski2018emergence}
K.~N. Durski, A.~M. McCollum, Y.~Nakazawa, B.~W. Petersen, M.~G. Reynolds, S.~Briand, M.~H. Djingarey, V.~Olson, I.~K. Damon, and A.~Khalakdina, ``Emergence of monkeypox—west and central africa, 1970--2017,'' \emph{Morbidity and mortality weekly report}, vol.~67, no.~10, p. 306, 2018.

\bibitem{mitja2023monkeypox}
O.~Mitj{\`a}, D.~Ogoina, B.~K. Titanji, C.~Galvan, J.-J. Muyembe, M.~Marks, and C.~M. Orkin, ``Monkeypox,'' \emph{The Lancet}, vol. 401, no. 10370, pp. 60--74, 2023.

\bibitem{who}
``{M}pox (monkeypox) --- who.int,'' \url{https://www.who.int/news-room/fact-sheets/detail/monkeypox}, 2023, [Accessed 03-08-2023].

\bibitem{li2006detection}
Y.~Li, V.~A. Olson, T.~Laue, M.~T. Laker, and I.~K. Damon, ``Detection of monkeypox virus with real-time pcr assays,'' \emph{Journal of Clinical Virology}, vol.~36, no.~3, pp. 194--203, 2006.

\bibitem{shafaati2022monkeypox}
M.~Shafaati and M.~Zandi, ``Monkeypox virus neurological manifestations in comparison to other orthopoxviruses,'' \emph{Travel Medicine and Infectious Disease}, vol.~49, p. 102414, 2022.

\bibitem{affonso2017deep}
C.~Affonso, A.~L.~D. Rossi, F.~H.~A. Vieira, A.~C.~P. de~Leon~Ferreira \emph{et~al.}, ``Deep learning for biological image classification,'' \emph{Expert systems with applications}, vol.~85, pp. 114--122, 2017.

\bibitem{shen2017deep}
D.~Shen, G.~Wu, and H.-I. Suk, ``Deep learning in medical image analysis,'' \emph{Annual review of biomedical engineering}, vol.~19, pp. 221--248, 2017.

\bibitem{rincy2020ensemble}
T.~N. Rincy and R.~Gupta, ``Ensemble learning techniques and its efficiency in machine learning: A survey,'' in \emph{2nd international conference on data, engineering and applications (IDEA)}.\hskip 1em plus 0.5em minus 0.4em\relax IEEE, 2020, pp. 1--6.

\bibitem{yu2022transfer}
X.~Yu, J.~Wang, Q.-Q. Hong, R.~Teku, S.-H. Wang, and Y.-D. Zhang, ``Transfer learning for medical images analyses: A survey,'' \emph{Neurocomputing}, vol. 489, pp. 230--254, 2022.

\bibitem{hu2018squeeze}
J.~Hu, L.~Shen, and G.~Sun, ``Squeeze-and-excitation networks,'' in \emph{Proceedings of the IEEE conference on computer vision and pattern recognition}, 2018, pp. 7132--7141.

\bibitem{almufareh2023transfer}
M.~F. Almufareh, S.~Tehsin, M.~Humayun, and S.~Kausar, ``A transfer learning approach for clinical detection support of monkeypox skin lesions,'' \emph{Diagnostics}, vol.~13, no.~8, p. 1503, 2023.

\bibitem{ali2022monkeypox}
S.~N. Ali, M.~T. Ahmed, J.~Paul, T.~Jahan, S.~Sani, N.~Noor, and T.~Hasan, ``Monkeypox skin lesion detection using deep learning models: A feasibility study,'' \emph{arXiv preprint arXiv:2207.03342}, 2022.

\bibitem{haque2023ensemble}
R.~Haque, A.~Sultana, and P.~Haque, ``Ensemble of fine-tuned deep learning models for monkeypox detection: A comparative study,'' in \emph{2023 4th International Conference for Emerging Technology (INCET)}.\hskip 1em plus 0.5em minus 0.4em\relax IEEE, 2023, pp. 1--8.

\bibitem{sahin2022human}
V.~H. Sahin, I.~Oztel, and G.~Yolcu~Oztel, ``Human monkeypox classification from skin lesion images with deep pre-trained network using mobile application,'' \emph{Journal of Medical Systems}, vol.~46, no.~11, p.~79, 2022.

\bibitem{haque2022classification}
M.~E. Haque, M.~R. Ahmed, R.~S. Nila, and S.~Islam, ``Classification of human monkeypox disease using deep learning models and attention mechanisms,'' \emph{arXiv preprint arXiv:2211.15459}, 2022.

\bibitem{irmak2022monkeypox}
M.~C. Irmak, T.~Aydin, and M.~Ya{\u{g}}ano{\u{g}}lu, ``Monkeypox skin lesion detection with mobilenetv2 and vggnet models,'' in \emph{2022 medical technologies congress (TIPTEKNO)}.\hskip 1em plus 0.5em minus 0.4em\relax IEEE, 2022, pp. 1--4.

\bibitem{altun2023monkeypox}
M.~Altun, H.~G{\"u}r{\"u}ler, O.~{\"O}zkaraca, F.~Khan, J.~Khan, and Y.~Lee, ``Monkeypox detection using cnn with transfer learning,'' \emph{Sensors}, vol.~23, no.~4, p. 1783, 2023.

\bibitem{nematzadeh2015comparative}
Z.~Nematzadeh, R.~Ibrahim, and A.~Selamat, ``Comparative studies on breast cancer classifications with k-fold cross validations using machine learning techniques,'' in \emph{2015 10th Asian control conference (ASCC)}.\hskip 1em plus 0.5em minus 0.4em\relax IEEE, 2015, pp. 1--6.

\bibitem{he2016deep}
K.~He, X.~Zhang, S.~Ren, and J.~Sun, ``Deep residual learning for image recognition,'' in \emph{Proceedings of the IEEE conference on computer vision and pattern recognition}, 2016, pp. 770--778.

\bibitem{omalley2019kerastuner}
T.~O'Malley, E.~Bursztein, J.~Long, F.~Chollet, H.~Jin, L.~Invernizzi \emph{et~al.}, ``Kerastuner,'' \url{https://github.com/keras-team/keras-tuner}, 2019.

\bibitem{liu2023convolutional}
R.~Liu, ``Convolutional siamese network-based few-shot learning for monkeypox detection under data scarcity,'' in \emph{Second International Conference on Biological Engineering and Medical Science (ICBioMed 2022)}, vol. 12611.\hskip 1em plus 0.5em minus 0.4em\relax SPIE, 2023, pp. 1459--1465.

\bibitem{pramanik2023monkeypox}
R.~Pramanik, B.~Banerjee, G.~Efimenko, D.~Kaplun, and R.~Sarkar, ``Monkeypox detection from skin lesion images using an amalgamation of cnn models aided with beta function-based normalization scheme,'' \emph{Plos one}, vol.~18, no.~4, p. e0281815, 2023.

\end{thebibliography}

\end{document}